\documentclass[prb,aps,twocolumn,amsmath,amssymb,floatfix,showpacs,
superscriptaddress]{revtex4}

\usepackage[dvips]{graphics}
\usepackage{color}
\definecolor{dgreen}{rgb}{0,0.6,0}

\begin{document}

\title{\textcolor{dgreen}{Magneto-transport in mesoscopic rings and 
cylinders: Effects of electron-electron interaction and spin-orbit coupling}}

\author{Santanu K. Maiti}

\email{santanu@post.tau.ac.il}

\affiliation{School of Chemistry, Tel Aviv University, Ramat-Aviv,
Tel Aviv-69978, Israel}

\begin{abstract}

We undertake an in-depth analysis of the magneto-transport properties in 
mesoscopic single-channel rings and multi-channel cylinders within a 
tight-binding formalism. The main focus of this review is to illustrate 
how the long standing anomalies between the calculated and measured current 
amplitudes carried by a small conducting ring upon the application of a 
magnetic flux $\phi$ can be removed. We discuss two different cases. First,
we examine the combined effect of second-neighbor hopping integral and 
Hubbard correlation on the enhancement of persistent current in presence
of disorder. A significant change in current amplitude is observed compared 
to the traditional nearest-neighbor hopping model and the current amplitude 
becomes quite comparable to experimental realizations. In the other case
we verify that in presence of spin-orbit interaction a considerable 
enhancement of persistent current amplitude takes place, and the current
amplitude in a disordered ring becomes almost comparable to that of an
ordered one. In addition to these, we also present the detailed band 
structures and some other related issues to get a complete picture of
the phenomena at the microscopic level.

\end{abstract}


\maketitle

\section{Introduction}

The study of magneto-transport properties in low-dimensional systems is 
always an interesting problem in mesoscopic physics. This is relatively a 
new branch of condensed matter physics which deals with systems whose 
dimensions are intermediate between the microscopic and macroscopic length 
scales~\cite{imrybook,datta1,datta2,metalidis,weinmann}. In contrast to the 
macroscopic objects where we generally use the laws of classical mechanics, 
the meso-scale systems are treated quantum mechanically since in this region 
fluctuations play the very crucial role. Over the last many years 
low-dimensional model quantum systems have been the objects of intense 
research, both in theory and in experiments, mainly due to the fact that
these simple looking systems are prospective candidates for nano devices
in electronic as well as spintronic engineering~\cite{aharon08,ahar2,
ore1,ore2,yey,torio,buks96,fuhrer07,kob,popp03,foldi08,berc04,berc2,
vidal1,vidal2,san10,san11,san12}. 

Several striking spectral properties are also exhibited by such systems 
owing to the quantum interference which is specially observed 
in quantum geometries with closed loop structures. The existence of 
dissipationless current in a mesoscopic metallic ring threaded by an 
Aharonov-Bohm (AB) flux $\phi$ is a direct consequence of quantum phase 
coherence. In this new quantum regime, two important aspects appear at low
temperatures. The first one is that the phase coherence length $L_{\phi}$
i.e., the length scale over which an electron maintains its phase memory,
increases significantly with the lowering of temperature and becomes 
comparable to the system size $L$. The other one is that the energy levels
of such small finite size systems are discrete. These two are the most 
essential criteria for the appearance of persistent charge current in a 
small metallic ring/cylinder due to the application of an external magnetic
flux $\phi$. In the pioneering work of B\"{u}ttiker, Imry and 
Landauer~\cite{butt}, the appearance of persistent current in metallic rings 
has been explored. Later, many excellent experiments~\cite{levy,chand,
mailly,jari,deb,reu} have been carried out in several ring and cylindrical 
geometries to reveal the actual mechanisms of persistent current. Though 
much efforts have been paid to study persistent current both 
theoretically~\cite{cheu1,cheu2,peeters1,peeters2,peeters3,mont,mont1,alts,
von,schm,ambe,abra,bouz,giam,yu,belu,ore,xiao1,xiao2,san1,san2,san3,san8}
as well as experimentally~\cite{levy,chand,mailly,jari,deb,reu}, yet
several anomalies still exist between the theory and experiment, and the
full knowledge about it in this scale is not well established even today.

The results of the single loop experiments are significantly different
from those for the ensemble of isolated loops. Persistent currents with
expected $\phi_0$ periodicity have been observed in isolated single Au
rings~\cite{chand} and in a GaAs-AlGaAs
ring~\cite{mailly}. Levy {\em et al.}~\cite{levy} found oscillations with
period $\phi_0/2$ rather than $\phi_0$ in an ensemble of $10^7$ independent
Cu rings. Similar $\phi_0/2$ oscillations were also reported for an ensemble
of disconnected $10^5$ Ag rings~\cite{deb} as well as for an array of $10^5$
isolated GaAs-AlGaAs rings~\cite{reu}. In a recent experiment, Jariwala
{\em et al.}~\cite{jari} obtained both $\phi_0$ and $\phi_0/2$ periodic
persistent currents in an array of thirty diffusive mesoscopic Au rings.
Except for the case of the nearly ballistic GaAs-AlGaAs ring~\cite{mailly},
all the measured currents are in general one or two orders of magnitude
larger than those expected from the theory.

Free electron theory predicts that at absolute zero temperature ($T=0\,$K), 
an ordered one-dimensional ($1$D) metallic ring threaded by magnetic flux 
$\phi$ 
supports persistent current with maximum amplitude $I_0=ev_F/L$, where 
$v_F$ is the Fermi velocity and $L$ is the circumference of the ring. 
Metals are intrinsically disordered which tends to decrease the persistent 
current, and the calculations show that the disorder-averaged current 
$\langle I \rangle$ crucially depends on the choice of the 
ensemble~\cite{cheu2,mont,mont1}. The magnitude of the current 
$\langle I^2\rangle^{1/2}$ is however insensitive to the averaging issues, 
and is of the order of $I_0 l/L$, $l$ being the elastic mean free path 
of the electrons. This expression remains valid even if one takes into 
account the finite width of the ring by adding contributions from the 
transverse channels, since disorder leads to a compensation between the 
channels~\cite{cheu2,mont}. However, the measurements on an ensemble 
of $10^7$ Cu rings~\cite{levy} reported a diamagnetic persistent current 
of average amplitude $3 \times 10^{-3}$ $ev_F/L$ with half a flux-quantum 
periodicity. Such $\phi_0/2$ oscillations with diamagnetic response were 
also found in other persistent current experiments consisting of ensemble 
of isolated rings~\cite{deb,reu}.

Measurements on single isolated mesoscopic rings on the other hand
detected $\phi_0$-periodic persistent currents with amplitudes
of the order of $I_0 \sim ev_F/L$, (closed to the value for an ordered ring).
Theory and experiment~\cite{mailly} seem to agree only when {\em disorder is
weak}. In another recent nice experiment Bluhm {\em et al.}~\cite{blu} have
measured the magnetic response of $33$ individual cold mesoscopic gold
rings, one ring at a time, using a scanning SQUID technique. They have
measured $h/e$ component and predicted that the measured current amplitude
agrees quite well with theory~\cite{cheu1} in a single ballistic
ring~\cite{mailly} and an ensemble of $16$ nearly ballistic
rings~\cite{raba}. However, the
amplitudes of the currents in single-isolated-diffusive
gold rings~\cite{chand} were two orders of magnitude larger than the
theoretical estimates. This discrepancy initiated intense theoretical
activity, and it is generally believed that the electron-electron
correlation plays an important role in the disordered diffusive
rings~\cite{abra,bouz,giam}.
An explanation based on the perturbative calculation in presence of
interaction and disorder has been proposed and it seems to give a
quantitative estimate closer to the experimental results, but still it
is less than the measured currents by an order of magnitude, and the
interaction parameter used in the theory is not well understood
physically. 

To remove the controversies regarding the persistent current amplitude 
between theoretical and experimental verifications we can proceed in two 
different ways. In literature almost all the theoretical results have been 
done based on a tight-binding (TB) framework within the {\em nearest-neighbor 
hopping} (NNH) approximation. It has been shown that in the NNH model 
electronic correlation provides a small enhancement of current amplitude 
in disordered materials i.e., a weak delocalizing effect is observed in 
presence of electron-electron (e-e) interaction. As a first attempt, we 
modify the traditional NNH model by incorporating the effects of higher 
order hopping integrals, at least second-neighbor hopping (SNH), in 
addition to the NNH integral. It is also quite physical since electrons 
have some finite probabilities to hop from one site to other sites apart 
from nearest-neighbor with reduced strengths. We will show that the 
inclusion of higher order hopping integrals gives significant enhancement 
of current amplitude and it reaches quite closer to the current amplitude 
of ordered systems. This is one approach. In the other way we examine 
that in presence of spin-orbit (SO) interaction a considerable enhancement
of persistent current amplitude takes place, and the current amplitude in
the disordered ring is almost comparable to that of an ordered ring.
The spin-orbit fields in a solid are called the Rashba spin-orbit 
interaction (RSOI) or the Dresselhaus spin-orbit interaction (DSOI) depending
on whether the electric field originates from a structural inversion
asymmetry or the bulk inversion asymmetry respectively~\cite{meier}.
Quantum rings formed at the interface of two semiconducting materials
are ideal candidates where the interplay of the two kinds of SOI might
be observed. A quantum ring in a heterojunction is realized when a two
dimensional gas of electrons is trapped in a quantum well due to the
{\it band offset} at the interface of two different semiconducting
materials. This {\it band offset} creates an electric field which may
be described by a potential gradient normal to the interface~\cite{premper}.
The potential at the interface is thus asymmetric, leading to the presence
of a RSOI. On the other hand, at such interfaces, the bulk inversion
symmetry is naturally broken.

The other important controversy comes for the determination of the sign
of low-field currents and still it is an unresolved issue between
theoretical and experimental results. In an experiment on persistent
current Levy {\em et al.}~\cite{levy} have shown diamagnetic nature for
the measured currents at low-field limit. While, in other experiment
Chandrasekhar {\em et al.}~\cite{chand} have obtained paramagnetic response
near zero field limit. Jariwala {\em et al.}~\cite{jari} have predicted
diamagnetic persistent current in their experiment and similar diamagnetic
response in the vicinity of zero field limit were also supported in an
experiment done by Deblock~\cite{deb} {\em et al.} on Ag rings. Yu and
Fowler~\cite{yu} have shown both diamagnetic and paramagnetic responses
in mesoscopic Hubbard rings. Though in a theoretical work Cheung {\em et
al.}~\cite{cheu2} have predicted that the direction of current is random
depending on the total number of electrons in the system and the specific
realization of the random potentials. Hence, prediction of the sign of
low-field currents is still an open challenge and further studies on
persistent current in mesoscopic systems are needed to remove the existing
controversies.

In the present review we address several important issues of 
magneto-transport in single-channel mesoscopic rings and multi-channel 
mesoscopic cylinders which are quite challenging from the standpoint of 
theoretical as well as experimental research. A brief outline of the 
presentation is as follows.

First, we address magnetic response in mesoscopic Hubbard rings threaded 
by AB flux $\phi$. We try to propose an idea to remove the unexpected 
discrepancy between the calculated and measured current amplitudes by 
incorporating the effect of second-neighbor hopping (SNH) in addition to 
the traditional nearest-neighbor hopping (NNH) integral in the 
tight-binding Hamiltonian. Using a generalized Hartree-Fock (HF)
approximation~\cite{kato,kam,san4,san5,san6}, we numerically compute 
persistent current ($I$), Drude weight ($D$) and low-field magnetic 
susceptibility ($\chi$) as functions of AB flux $\phi$, total number 
of electrons $N_e$ and system size $N$. With this (HF) approach one 
can study magnetic response in a much larger system since here a many-body 
Hamiltonian is decoupled into two effective one-body Hamiltonians. One
is associated with up spin electrons and other is related to down spin
electrons. But the point is that, the results calculated using generalized
HF mean-field theory may deviate from exact results with the reduction of
dimensionality. So we should take care about the mean-field calculation,
specially, in $1$D systems. To trust our predictions, here we also we 
make a comparative study between the results obtained from mean-field 
theory and exactly diagonalizing the full many-body Hamiltonian. The later 
approach where a complete many-body Hamiltonian is diagonalized to get 
energy eigenvalues is not suitable to study magnetic response in larger 
systems since the size of the matrices increases very sharply with the 
total number of up and down spin electrons.

Next, we explore the behavior of persistent current in an interacting 
mesoscopic ring with finite width threaded by an Aharonov-Bohm flux $\phi$. 
For this cylindrical system we also see that the inclusion of higher order
hopping integrals leads to a possibility of getting enhanced persistent 
current and the current is quite comparable to the ordered one. Our results 
can be utilized to study magnetic response in any interacting mesoscopic 
system.

Finally, in the last part, we focus our attention on the behavior of 
persistent current in a one-dimensional mesoscopic ring threaded by a 
magnetic flux in presence of the Rashba and Dresselhaus SO interactions. 
Here, the effect of electron-electron interaction is neglected. We show 
that the presence of the SO interaction leads to a significant enhancement 
of persistent current~\cite{san7}. In addition to these, we also describe 
very briefly the energy band structures and the oscillations of persistent 
current as the RSOI is varied to make the present communication a self 
contained study.

Throughout the review we perform all the essential features of 
magneto-transport at absolute zero temperature and set $c=e=h=1$ for 
numerical calculations.

\section{A Hubbard ring in absence of SO interactions}

In this section we describe the magneto-transport properties in a 
single-channel $1$D mesoscopic ring in the presence of on-site Coulomb
interaction. The effect of SO interaction is not taken into account.

\subsection{Model and theoretical formulation}

\begin{figure}[ht]
{\centering \resizebox*{4.3cm}{2.5cm}{\includegraphics{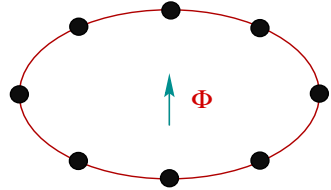}}\par}
\caption{(Color online). Schematic view of a $1$D mesoscopic ring
penetrated by a magnetic flux $\phi$. The filled black circles
correspond to the positions of the atomic sites.}
\label{ring}
\end{figure}

We start by referring to Fig.~\ref{ring}, where a normal metal ring
is threaded by a magnetic flux $\phi$.
To describe the system we use a tight-binding framework. For a $N$-site
ring, penetrated by a magnetic flux $\phi$ (measured in unit of the
elementary flux quantum $\phi_0=ch/e$), the tight-binding Hamiltonian
in Wannier basis looks in the form,
\begin{eqnarray}
H_R & = &\sum_{i,\sigma}\epsilon_{i\sigma} c_{i\sigma}^{\dagger} c_{i\sigma}
+\sum_{ij,\sigma} t \left[e^{i\theta} c_{i\sigma}^{\dagger} 
c_{j\sigma} + h.c. \right] + \nonumber \\
&  & \sum_{ik,\sigma} t_1 \left[e^{i\theta_1} c_{i\sigma}^{\dagger} 
c_{k\sigma} + h.c. \right] + \sum_i U c_{i\uparrow}^{\dagger}c_{i\uparrow} 
c_{i\downarrow}^{\dagger} c_{i\downarrow}
\label{equ1}
\end{eqnarray}
where, $\epsilon_{i\sigma}$ is the on-site energy of an electron at
the site $i$ of spin $\sigma$ ($\uparrow,\downarrow$). The variable $t$
corresponds to the nearest-neighbor ($j=i\pm1$) hopping strength, while
$t_1$ gives the second-neighbor ($k=i\pm2$) hopping integral.
$\theta=2\pi\phi/N$ and $\theta_1=4\pi\phi/N$ are the phase factors
associated with the hopping of an electron from one site to its neighboring
site and next-neighboring site, respectively. $c_{i\sigma}^{\dagger}$ and
$c_{i\sigma}$ are the creation and annihilation operators, respectively,
of an electron at the site $i$ with spin $\sigma$. $U$ is the strength of
on-site Hubbard interaction.

\vskip 0.2cm
\noindent
{\bf Decoupling of the interacting Hamiltonian:}
In order to determine the energy eigenvalues of the interacting model
quantum system described by the tight-binding Hamiltonian given in
Eq.~\ref{equ1}, first we decouple the interacting Hamiltonian using
generalized Hartree-Fock approach, the so-called mean field approximation.
In this approach, the full Hamiltonian is completely decoupled into two
parts. One is associated with the up-spin electrons, while the other is
related to the down-spin electrons with their modified site energies.
For up and down spin Hamiltonians, the modified site energies are
expressed in the form,
$\epsilon_{i\uparrow}^{\prime}=\epsilon_{i\uparrow} + U \langle 
n_{i\downarrow} \rangle$ and
$\epsilon_{i\downarrow}^{\prime}=\epsilon_{i\downarrow} + U \langle 
n_{i\uparrow} \rangle$,
where $n_{i\sigma}=c_{i\sigma}^{\dagger} c_{i\sigma}$ is the number
operator. With these site energies, the full Hamiltonian (Eq.~\ref{equ1})
can be written in the decoupled form as,
\begin{eqnarray}
H_R &=&\sum_i \epsilon_{i\uparrow}^{\prime} n_{i\uparrow} + \sum_{ij} t 
\left[e^{i\theta} c_{i\uparrow}^{\dagger} c_{j\uparrow} + e^{-i\theta} 
c_{j\uparrow}^{\dagger} c_{i\uparrow}\right] \nonumber \\
& + & \sum_{ik} t_1 \left[e^{i\theta_1} c_{i\uparrow}^{\dagger} c_{k\uparrow}
+ e^{-i\theta_1} c_{k\uparrow}^{\dagger} c_{i\uparrow}\right] \nonumber \\
& + & \sum_i \epsilon_{i\downarrow}^{\prime} n_{i\downarrow} + \sum_{ij} 
t \left[e^{i\theta} c_{i\downarrow}^{\dagger} c_{j\downarrow}
+ e^{-i\theta} c_{j\downarrow}^{\dagger} c_{i\downarrow}\right] \nonumber \\
& + & \sum_{ik} t_1 \left[e^{i\theta_1} c_{i\downarrow}^{\dagger} 
c_{k\downarrow} + e^{-i\theta_1} c_{k\downarrow}^{\dagger} c_{i\downarrow}
\right] \nonumber \\
& - & \sum_i U \langle n_{i\uparrow} \rangle \langle n_{i\downarrow} 
\rangle \nonumber \\
&=& H_{\uparrow}+H_{\downarrow}-\sum_i U \langle n_{i\uparrow} 
\rangle \langle n_{i\downarrow} \rangle
\label{equ4} 
\end{eqnarray}
where, $H_{\uparrow}$ and $H_{\downarrow}$ correspond to the effective
tight-binding Hamiltonians for the up and down spin electrons, respectively.
The last term is a constant term which provides an energy shift in the
total energy.

\vskip 0.2cm
\noindent
{\bf Self consistent procedure:}
With these decoupled Hamiltonians ($H_{\uparrow}$ and $H_{\downarrow}$)
of up and down spin electrons, we start our self consistent procedure
considering initial guess values of $\langle n_{i\uparrow} \rangle$ and
$\langle n_{i\downarrow} \rangle$. For these initial set of values of
$\langle n_{i\uparrow} \rangle$ and $\langle n_{i\downarrow} \rangle$,
we numerically diagonalize the up and down spin Hamiltonians. Then we
calculate a new set of values of $\langle n_{i\uparrow} \rangle$ and
$\langle n_{i\downarrow} \rangle$. These steps are repeated until a self
consistent solution is achieved.

\vskip 0.2cm
\noindent
{\bf Calculation of ground state energy:}
Using the self consistent solution, the ground state energy $E_0$ for a 
particular filling at absolute zero temperature ($T=0\,$K) can be 
determined by taking the sum of individual states up to Fermi energy 
($E_F$) for both up and down spins. Thus, we can write the final form of 
ground state energy as,
\begin{equation}
E_0=\sum_n E_{n\uparrow} + \sum_n E_{n\downarrow}- \sum_i U \langle
n_{i\uparrow} \rangle \langle n_{i\downarrow} \rangle
\label{equ5} 
\end{equation}
where, the index $n$ runs for the states upto the Fermi level.
$E_{n\uparrow}$ ($E_{n\downarrow}$) is the single particle energy 
eigenvalue for $n$-th eigenstate obtained by diagonalizing the 
Hamiltonian $H_{\uparrow}$ ($H_{\downarrow}$).

\vskip 0.2cm
\noindent
{\bf Calculation of persistent current:}
At absolute zero temperature, total persistent current of the system
is obtained from the expression,
\begin{equation}
I(\phi)=-c\frac{\partial E_0(\phi)}{\partial \phi}
\end{equation}
where, $E_0(\phi)$ is the ground state energy.

\vskip 0.2cm
\noindent
{\bf Calculation of Drude weight:}
The Drude weight for the ring can be calculated through the relation,
\begin{equation}
D=\left . \frac{N}{4\pi^2} \left(\frac{\partial{^2E_0(\phi)}}
{\partial{\phi}^{2}}\right) \right|_{\phi \rightarrow 0}
\label{equ7}
\end{equation}
where, $N$ gives total number of atomic sites in the ring. Kohn~\cite{kohn}
has shown that for an insulating system $D$ decays exponentially to zero,
while it becomes finite for a conducting system.

\vskip 0.2cm
\noindent
{\bf Determination of low-field magnetic susceptibility:}
The general expression of magnetic susceptibility $\chi$ at any flux
$\phi$ is written in the form,
\begin{equation}
\chi(\phi)=\frac{N^3}{16\pi^2}\left(\frac{\partial I(\phi)}{\partial \phi}
\right).
\label{equ8}
\end{equation}
Evaluating the sign of $\chi(\phi)$ we can able to predict whether the
current is paramagnetic or diamagnetic in nature. Here we will determine
$\chi(\phi)$ only in the limit $\phi \rightarrow 0$, since we are
interested to know the magnetic response in the low-field limit.

\subsection{Numerical results and discussion}

In this sub-section throughout our numerical work we set the 
nearest-neighbor hopping strength $t=-1$ and second-neighbor hopping 
strength $t_1=-0.7$. Energy scale is measured in unit of $t$.

\subsubsection{Perfect Hubbard Rings Described with NNH Integral}

For perfect rings we choose $\epsilon_{i \uparrow}=
\epsilon_{i \downarrow}=0$ for all $i$ and since here we consider the
rings with only NNH integral, the second-neighbor hopping strength $t_1$
is fixed at zero.

\begin{figure}[ht]
{\centering \resizebox*{7.75cm}{8cm}{\includegraphics{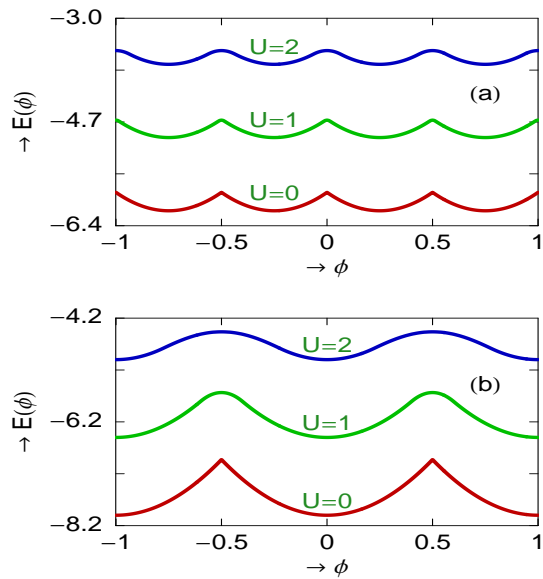}}\par}
\caption{(Color online). Ground state energy levels as a function of flux
$\phi$ for some typical mesoscopic rings in half-filled case. The red,
green and blue curves correspond to $U=0$, $1$ and $2$, respectively.
(a) $N=5$ and (b) $N=6$.}
\label{ringenergy}
\end{figure}

\vskip 0.2cm
\noindent
{\bf Energy-flux characteristics:}
To explain the relevant features of magnetic response we begin with
the energy-flux characteristics. As illustrative examples, in
Fig.~\ref{ringenergy} we plot the ground state energy levels as a
function of magnetic flux $\phi$ for some typical mesoscopic rings in
the half-filled case, where (a) and (b) correspond to $N=5$ and $6$,
respectively. The red curves represent the energy levels for the
non-interacting ($U=0$) rings, while the green and blue lines correspond
to the energy levels for the interacting rings where the electronic
correlation strength $U$ is
fixed to $1$ and $2$, respectively. From the spectra it is observed that
the ground state energy level shifts towards the positive energy and it
becomes much flatter with the increase of the correlation strength $U$.
Both for the two different ring sizes ($N=5$ and $6$) the ground state
energy levels vary periodically with AB flux $\phi$, but a significant
difference is observed in their periodicities depending on the oddness
and evenness of the ring size $N$. For $N=6$ (even), the energy levels
show conventional $\phi_0$ ($=1$, in our chosen unit system $c=e=h=1$)
flux-quantum periodicity. On the other hand, the period becomes half i.e.,
$\phi_0/2$ for $N=5$ (odd). This $\phi_0/2$ periodicity disappears as long
as the filling is considered away from the half-filling. At the same time,
it also vanishes if impurities are introduced in the system, even if the
ring is half-filled with odd $N$. Therefore, $\phi_0/2$ periodicity is a
special feature for odd half-filled perfect rings irrespective of the
Hubbard strength $U$, while for all other cases traditional $\phi_0$
periodicity is obtained.

\begin{figure}[ht]
{\centering \resizebox*{7.75cm}{8cm}{\includegraphics{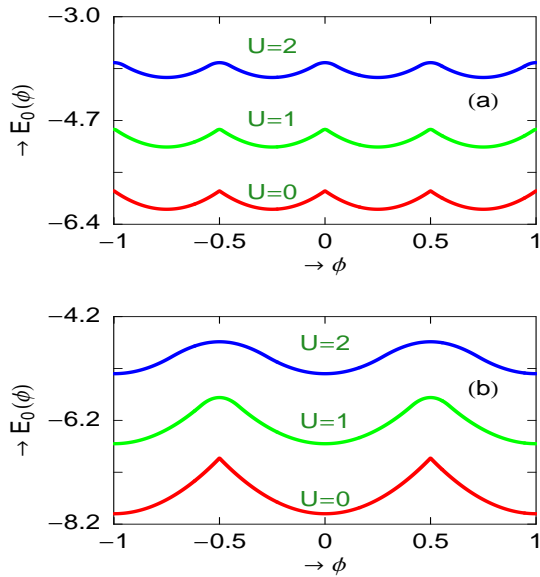}}\par}
\caption{(Color online). Ground state energy levels as a function of flux
$\phi$ for some typical mesoscopic rings in half-filled case, where
eigenenergies are determined through exact diagonalization of the full
many-body Hamiltonian. The red, green and blue curves correspond to $U=0$,
$1$ and $2$, respectively. (a) $N=5$ and (b) $N=6$.}
\label{exactenergy}
\end{figure}

To judge the accuracy of the mean-field calculations
in our ring geometry, in Fig.~\ref{exactenergy} we show the variation of
lowest energy levels where the eigenenergies are determined through exact
diagonalization of the full many-body Hamiltonian for the identical rings
as given in Fig.~\ref{ringenergy}, considering the same parameter values.
Comparing the results presented in Figs.~\ref{ringenergy} and
\ref{exactenergy}, we see that the mean-field results agree very well
with the exact diagonalization method. Thus we can safely use mean-field
approach to study magnetic response in our geometry.

\begin{figure}[ht]
{\centering \resizebox*{7.75cm}{8cm}{\includegraphics{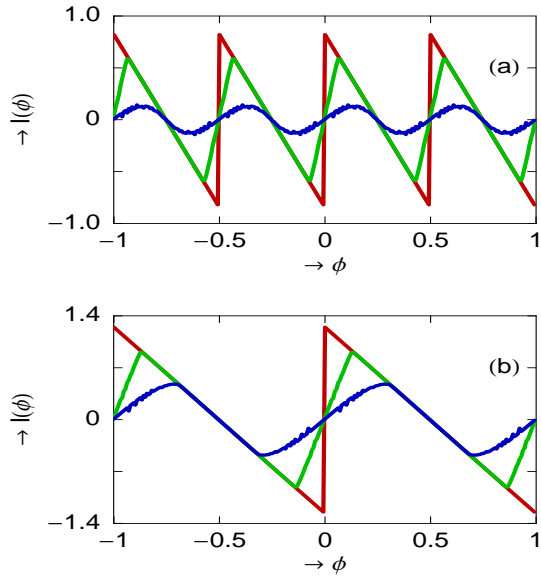}}\par}
\caption{(Color online). Persistent current as a function of flux $\phi$
for single-channel mesoscopic rings in half-filled case. (a) $N=15$. The
red, green and blue curves correspond to $U=0$, $1.5$ and $2$, respectively.
(b) $N=20$. The red, green and blue curves correspond to $U=0$, $1$ and
$1.5$, respectively.}
\label{ringcurr}
\end{figure}

\vskip 0.2cm
\noindent
{\bf Current-flux characteristics:}
Following the above energy-flux characteristics now we describe the
behavior of persistent current in mesoscopic Hubbard rings. As
representative examples, in Fig.~\ref{ringcurr} we display the variation
of persistent currents as a function of flux $\phi$ for some typical
single-channel mesoscopic rings in the half-filled case, where (a) and
(b) correspond to $N=15$ and $20$, respectively. The red, green and blue
curves in Fig.~\ref{ringcurr}(a) correspond to the currents for $U=0$,
$1.5$ and $2$, respectively, while these curves in Fig.~\ref{ringcurr}(b)
represent the currents for $U=0$, $1$ and $1.5$, respectively. In the
absence of any e-e interaction ($U=0$), persistent current shows
saw-tooth like nature as a function of flux $\phi$ with sharp transitions
at $n\phi_0/2$ (red line of Fig.~\ref{ringcurr}(a)) or $n\phi_0$ (red
line of Fig.~\ref{ringcurr}(b)), where $n$ being an integer, depending on
whether $N$ is odd or even. The saw-tooth like behavior disappears as
long as the electronic correlation is introduced into the system. This
is clearly observed from the green and blue curves of Fig.~\ref{ringcurr}.
Additionally, in the presence of $U$, the current amplitude gets
suppressed compared to the current amplitude in the non-interacting case,
and it decreases gradually with increasing $U$. This provides the lowering
of electron mobility with the rise of $U$ and the reason behind this can
be much better understood from our forthcoming discussion. Both for two
different rings with sizes $N=15$ (odd) and $20$ (even), persistent
currents vary periodically with AB flux $\phi$ showing different
periodicities, following the energy-flux characteristics. For $N=15$,
current shows $\phi_0/2$ flux-quantum periodicity, while for the other
case ($N=20$), current exhibits $\phi_0$ flux-quantum periodicity.

\vskip 0.2cm
\noindent
{\bf Variation of electronic mobility-Drude weight:}
To reveal the conducting properties of Hubbard rings, we study the
variation of Drude weight $D$ for these systems. Drude weight can be
calculated by using Eq.~\ref{equ7}. Finite value of $D$ predicts the
metallic phase, while for the insulating phase it drops exponentially
to zero~\cite{kohn}.

\begin{figure}[ht]
{\centering \resizebox*{7.75cm}{8cm}{\includegraphics{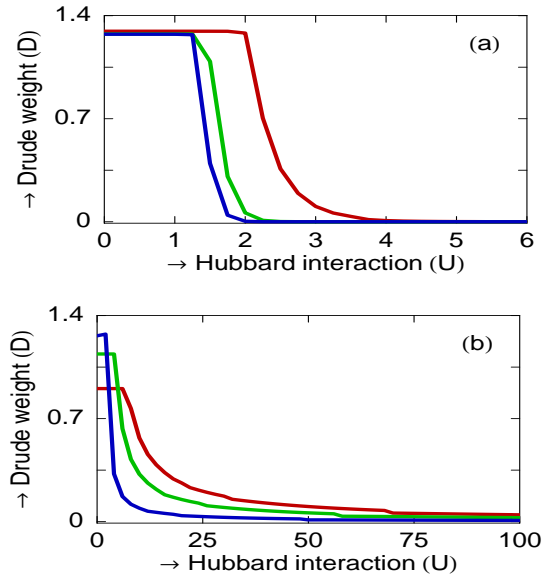}}\par}
\caption{(Color online). Drude weight as a function of Hubbard interaction
strength $U$ for single-channel mesoscopic rings. (a) Half-filled case.
The red, green and blue curves correspond to $N=10$, $30$ and $50$,
respectively. (b) Non-half-filled case with $N=20$. The red, green and
blue curves correspond to $N_e=10$, $14$ and $18$, respectively.}
\label{ringdrude}
\end{figure}

\begin{figure}[ht]
{\centering \resizebox*{7.75cm}{8cm}{\includegraphics{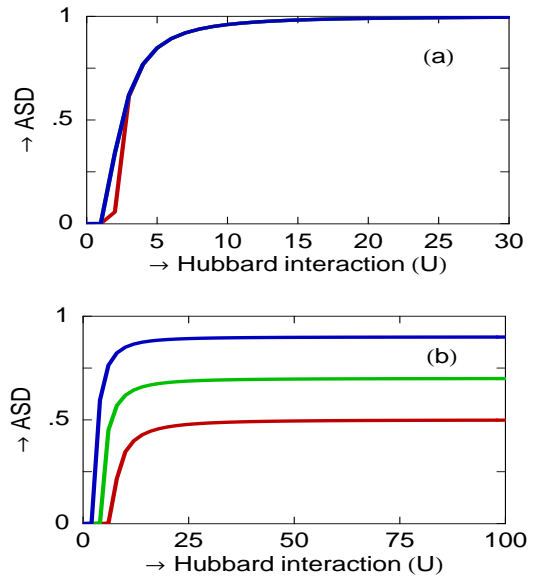}}\par}
\caption{(Color online). Average spin density (ASD) as a function of
Hubbard interaction strength $U$ for single-channel mesoscopic rings.
(a) Half-filled case. The red, green and blue curves correspond to $N=10$,
$30$ and $50$, respectively. (b) Non-half-filled case with $N=20$. The red,
green and blue curves correspond to $N_e=10$, $14$ and $18$, respectively.}
\label{ringspinden}
\end{figure}

As illustrative examples, in Fig.~\ref{ringdrude} we show the variation
of Drude weight $D$ as a function of electronic correlation strength $U$
for some typical single-channel Hubbard rings. In Fig.~\ref{ringdrude}(a)
the results are shown for three different half-filled rings, where the
red, green and blue lines correspond to the rings with $N=10$, $30$ and
$50$, respectively. From the curves it is evident that for smaller
values of $U$, the half-filled rings show finite value of $D$ which
reveals that they are in the metallic phase. On the other hand, $D$
drops sharply to zero when $U$ becomes high. Thus the rings become
insulating when $U$ is quite large. The results for the non-half filled
case are shown in Fig.~\ref{ringdrude}(b), where we fix the ring size
$N=20$ and vary the electron filling. The red, green and blue curves
represent $N_e=10$, $14$ and $18$, respectively, where $N_e$ gives the
total number of electrons in the ring. For these three choices of $N_e$,
the ring is always less than half-filled (since $N_e<N$) and the ring
is in the conducting phase irrespective of the correlation strength $U$.
Now we try to justify the dependence of the Hubbard strength $U$ on the
electronic mobility for these different fillings. To understand the
effect of $U$ on electron mobility here we measure a quantity called
`average spin density' (ASD) which is defined by the factor $\sum_i 
|(n_{i\uparrow}-n_{i\downarrow})|/N$. The integer $i$ is the site index
and it runs from $1$ to $N$. By calculating ASD we can estimate the
occupation probability of electrons in the ring and it supports us
to explain whether the ring lies in the metallic phase or in the
insulating one. For the rings those are below half-filled, ASD is always
less than unity irrespective of the value of $U$ as shown by the curves
in Fig.~\ref{ringspinden}(b). It reveals that for these systems, ground
state always supports an empty site and electron can move along the
ring avoiding double occupancy of two different spin electrons at any
site $i$ in the presence of e-e correlation which provides the metallic
phase ($D>0$). For a fixed ring size and a particular strength of $U$,
the ASD increases as the filling is increased towards half-filling which
is noticed by comparing the three different curves in
Fig.~\ref{ringspinden}(b). On the other hand, in the half-filled rings,
ASD is less than unity for small value of $U$, while it reaches to unity
when $U$ is large. This behavior is clearly shown by the curves given in
Fig.~\ref{ringspinden}(a), where the red, green and blue lines correspond
to ASDs for the half-filled rings with $N=10$, $30$ and $50$, respectively.
Thus, for low $U$ there is some finite probability of getting two opposite
spin electrons in a same site which allows electrons to move along the ring
and the metallic phase is obtained. But for large $U$, ASD reaches to
unity which means that each site is singly occupied either by an up or
down spin electron with probability $1$. In this case ground state
does not support any empty site and due to strong repulsive e-e
correlation one electron sitting in a site does not allow to come other
electron with opposite spin from the neighboring site which provides the
insulating phase ($D=0$). The situation is somewhat analogous to Mott
localization in one-dimensional infinite lattices.
In perfect Hubbard rings the conducting nature has been
studied exactly quite a long ago using the ansatz of Bethe by Shastry
and Sutherland~\cite{shastry}. They have calculated charge stiffness
constant ($D_c$) and have predicted that $D_c$ goes to zero as the system
approaches towards half-filling for any non-zero value of $U$. Our
numerical results clearly justify their findings.

\begin{figure}[ht]
{\centering \resizebox*{7.75cm}{8cm}{\includegraphics{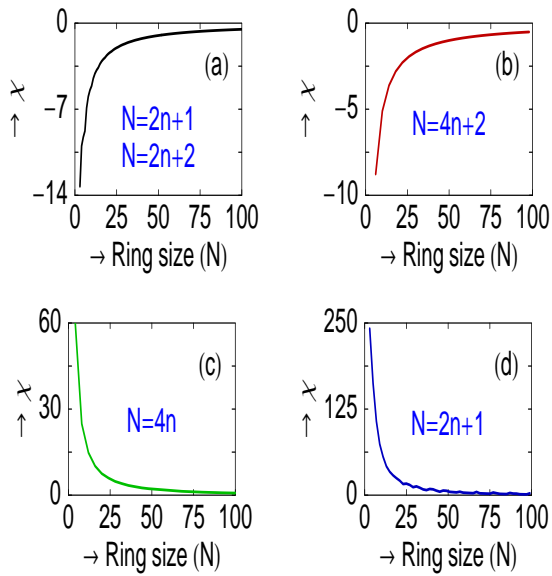}}\par}
\caption{(Color online). Low-field magnetic susceptibility as a function of
system size $N$ for single-channel mesoscopic rings in half-filled case.
(a) $U=0$. $N$ is an odd ($2n+1$) or an even ($2n+2$) number, where $n$ is
an integer. (b) $U=1$. $N$ is an even number obeying the relation $N=4n+2$.
(c) $U=1$. $N$ is an even number satisfying the relation $N=4n$. (d) $U=1$.
$N$ is an odd number following the relation $N=2n+1$.}
\label{ringsuscep}
\end{figure}

\vskip 0.2cm
\noindent
{\bf Low-field magnetic susceptibility:}
Now, we discuss the variation of low-field magnetic susceptibility which
can be calculated from Eq.~\ref{equ8} by setting $\phi \rightarrow 0$.
With the help of this parameter we can justify
whether the current is paramagnetic ($+$ve slope) or diamagnetic ($-$ve
slope) in nature. For our illustrative purposes, in Fig.~\ref{ringsuscep}
we show the variation of low-field magnetic susceptibility with system
size $N$ for some typical single-channel mesoscopic rings in the
half-filled case. Figure~\ref{ringsuscep}(a) correspond to the variation
of low-field magnetic susceptibility for the non-interacting ($U=0$) rings,
where the ring size can by anything i.e., either odd, following the
relation $N=2n+1$ ($n$ is an integer), or even, obeying the expression
$N=2n+2$. It is observed that both for odd and even $N$, low-field current
exhibits diamagnetic nature. The behavior of the low-field currents changes
significantly when the e-e interaction is taken into account. Depending
on the ring size $N$, the sign becomes $+$ve and $-$ve as shown by the
curves given in Figs.~\ref{ringsuscep}(b)-(d). For the interacting rings
where the relation $N=4n+2$ is satisfied, the low-field current becomes
diamagnetic (Fig.~\ref{ringsuscep}(b)). The sign becomes paramagnetic when
$N=4n$ (Fig.~\ref{ringsuscep}(c)) and $N=2n+1$ (Fig.~\ref{ringsuscep}(d)).
Thus, in brief, we say that for non-interacting half-filled rings low-field
current exhibits diamagnetic response irrespective of $N$ i.e., whether
$N$ is odd or even. For the interacting half-filled rings with odd $N$,
low-field current provides only the paramagnetic behavior, while for
even $N$, depending on the particular value of $N$, the response becomes
either diamagnetic or paramagnetic. These natures of low-field currents
change for the cases of other electron fillings. Hence, it can be
emphasized that the behavior of the low-field currents is highly sensitive
on the Hubbard correlation, electron filling, evenness and oddness of $N$,
etc. The behavior of zero-field magnetic susceptibility
in Hubbard rings has been studied extensively quite a long back using the
Bethe ansatz by Shiba~\cite{shiba}. In this work, he has studied magnetic
susceptibility per electron as functions of electron filling and Hubbard
correlation strength and provided several interesting results. From his
findings we can clearly justify our presented results.

\subsubsection{Disordered Hubbard Rings Described with NNH and SNH
Integrals}

Now, we explore the combined effect of electron-electron correlation
and second-neighbor hopping (SNH) integral on persistent current in
disordered mesoscopic rings.

To get a disordered ring, we choose site energies ($\epsilon_{i\uparrow}$
and $\epsilon_{i\downarrow}$) randomly from a ``Box" distribution function
of width $W$. As the site energies are chosen randomly it is needed to
consider the average over a large number of disordered configurations (from
the stand point of statistical average). Here, we determine the currents by
taking the average over $50$ random disordered configuration in each case
to achieve much accurate results.

\begin{figure}[ht]
{\centering \resizebox*{7.75cm}{8cm}{\includegraphics{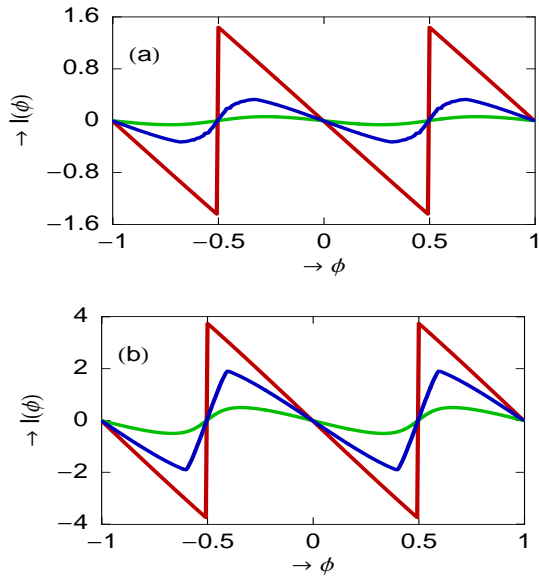}}\par}
\caption{(Color online). Persistent current as a function of flux $\phi$
for single-channel mesoscopic rings with $N=15$ considering $1/3$ electron
filling. (a) Rings with only NNH integral. The red line corresponds to the
ordered non-interacting ring, while the green and blue lines correspond
to the disordered ($W=2$) rings with $U=0$ and $0.5$, respectively. (b)
Rings with NNH and SNH integrals. The red line represents the ordered
non-interacting ring, whereas the green and blue line correspond to the
disordered ($W=2$) rings with $U=0$ and $1.5$, respectively.}
\label{ringcurrsnh}
\end{figure}

As illustrative examples, in Fig.~\ref{ringcurrsnh} we display the
variation of persistent currents for some single-channel mesoscopic
rings considering $1/3$ electron filling. In (a) the results are given
for the rings characterized by the NNH integral model. The red curve
represents the current for the ordered ($W=0$) non-interacting ($U=0$)
ring. It shows saw-tooth like nature with AB flux $\phi$ providing $\phi_0$
flux-quantum periodicity. The situation becomes completely different when
impurities are introduced in the ring as clearly seen by the other two
colored curves. The green curve represents the current for the case only
when impurities are considered but the effect of Hubbard interaction is
not taken into account. It varies continuously with $\phi$ and gets much
reduced amplitude, even an order of magnitude, compared to the perfect case.
This is due to the localization of the energy eigenstates in the presence
of impurity, which is the so-called Anderson localization. Hence, a large
difference exists between the current amplitudes of an ordered and
disordered non-interacting rings and it was the main controversial issue
among the theoretical and experimental predictions. Experimental results
suggest that the measured current amplitude is quite comparable to the
theoretically estimated current amplitude in a perfect system. To remove
this controversy, as a first attempt, we include the effect of Hubbard
interaction in the disordered ring described by the NNH model. The result
is shown by the blue curve where $U$ is fixed at $0.5$. It is observed
that the current amplitude gets increased compared to the non-interacting
disordered ring, though the increment is too small. Not only that the
enhancement can take place only for small values of $U$, while for large
enough $U$ the current amplitude rather decreases. This phenomenon can
be explained as follows. For the non-interacting disordered ring the
probability of getting two opposite spin electrons becomes higher at
the atomic sites where the site energies are lower than the other sites
since the electrons get pinned at the lower site energies to minimize
the ground state energy, and this pinning of electrons becomes increased
with the rise of impurity strength $W$. As a result the mobility of
electrons and hence the current amplitude gets reduced with the increase
of impurity strength $W$. Now, if we introduce electronic correlation in
the system then it tries to depin two opposite spin electrons those are
situated together due to the Coulomb repulsion. Therefore, the electronic
mobility is enhanced which provides quite larger current amplitude. But,
for large enough interaction strength, mobility of electrons gradually
decreases due to the strong repulsive interaction. Accordingly, the current
amplitude gradually decreases with $U$. So, in short, we can say that
within the nearest-neighbor hopping (NNH) model electron-electron
interaction does not provide any significant contribution to enhance
the current amplitude, and hence the controversy regarding the current
amplitude still persists.

To overcome this controversy, finally we make an attempt by incorporating
the effect of second-neighbor hopping (SNH) integral in addition to the
nearest-neighbor hopping (NNH) integral. With this modification a
significant change in current amplitude takes place which is clearly
observed from Fig.~\ref{ringcurrsnh}(b). The red curve refers
to the current for the perfect ($W=0$) non-interacting ($U=0$) ring
and it achieves much higher amplitude compared to the NNH model (see red
curve of Fig.~\ref{ringcurrsnh}(a)). This additional contribution
comes from the SNH integral since it allows electrons to hop further.
The main focus of this sub-section is to interpret the combined effect
of SNH integral and Hubbard correlation on the enhancement of persistent
current in disordered ring. To do this first we narrate the effect of
SNH integral in disordered non-interacting ring.
The nature of the current for this particular case is shown by the green
curve of Fig.~\ref{ringcurrsnh}(b). It shows that the current amplitude
gets reduced compared to the perfect case (red line), which is expected,
but the reduction of the current amplitude is very small than the NNH
integral model (see green curve of Fig.~\ref{ringcurrsnh}(a)). This is
due the fact that the SNH integral tries to delocalize the electronic
states, and therefore, the mobility of the electrons is enriched. The
situation becomes more interesting when we include the effect of Hubbard
interaction. The behavior of the current in the presence of interaction
is plotted by the blue curve of Fig.~\ref{ringcurrsnh}(b) where we fix
$U=1.5$. Very interestingly we see that the current amplitude is enhanced
moderately and quite comparable to that of the perfect ring. Therefore,
it can be predicted that the presence of SNH integral and Hubbard
interaction can provide a persistent current which may be comparable to the
measured current amplitudes. In the above analysis we consider the effect 
of only SNH integral in addition to the NNH model, and, illustrate how such
a higher order hopping integral leads to an important role on the enhancement
of current amplitude in presence of Hubbard correlation for disordered
rings. Instead of considering only the SNH integral we can also take
the contributions from all possible higher order hopping integrals with
reduced hopping strengths. Since the strengths of other higher order
hopping integrals are too small, the contributions from these factors
are reasonably small and they will not provide any significant change in
the current amplitude. 

\section{A Hubbard cylinder in absence of SO interactions}

In this section we extend our discussion for an interacting mesoscopic 
ring with finite width threaded by an AB flux $\phi$. Here also we ignore 
the effect of SO interaction on magneto-transport properties like the 
previous section.

\subsection{Model and the Hamiltonian}

\begin{figure}[ht]
{\centering \resizebox*{5cm}{3cm}{\includegraphics{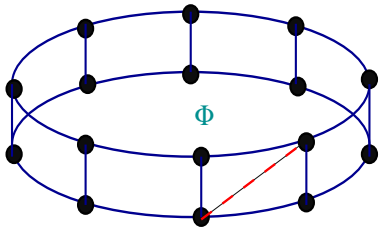}}\par}
\caption{(Color online). Schematic view of a $1$D mesoscopic cylinder
penetrated by a magnetic flux $\phi$. The red dashed line corresponds to
the second-neighbor hopping integral and the filled black circles represent
the positions of the atomic sites. A persistent current $I$ is established
in the cylinder.}
\label{cylinder}
\end{figure}

Let us start by referring to Fig.~\ref{cylinder}, where a small metallic
cylinder is threaded by a magnetic flux $\phi$. The filled black circles
correspond to the positions of the atomic sites in the cylinder. To predict
the size of a cylinder we use two parameters $N$ and $M$, where the $1$st
one ($N$) represents total number of atomic sites in each circular ring
and the other one ($M$) gives total number of identical circular rings.
For the description of our model quantum system we use a tight-binding
framework and in order to incorporate the effect of higher order
hopping integrals to the Hamiltonian here we consider second-neighbor
hopping (SNH) (shown by the red dashed line in Fig.~\ref{cylinder}) in
addition to the nearest-neighbor hopping (NNH) of electrons. Considering
both NNH and SNH integrals the TB Hamiltonian for the cylindrical system
in Wannier basis looks in the form,
\begin{eqnarray}
H_{\mbox{c}} & = &\sum_{i,j,\sigma}\epsilon_{i,j,\sigma} 
c_{i,j,\sigma}^{\dagger} c_{i,j,\sigma} + \sum_{i,j,\sigma} 
t_l \left[e^{i\theta_l} c_{i,j,\sigma}^{\dagger} c_{i,j+1,\sigma}
\right. \nonumber \\ 
& + & \left. h.c. \right] + \sum_{i,j,\sigma} t_d \left[e^{i\theta_d} 
c_{i,j,\sigma}^{\dagger} c_{i+1,j+1,\sigma} + h.c. \right] \nonumber \\
& + & \sum_{ij} U c_{i,j,\uparrow}^{\dagger}c_{i,j,\uparrow} 
c_{i,j,\downarrow}^{\dagger} c_{i,j,\downarrow}
\label{equcyl}
\end{eqnarray}
where, ($i,j$) represent the co-ordinate of a lattice site. The index
$i$ runs from $1$ to $M$, while the integer $j$ goes from $1$ to $N$.
$\epsilon_{i,j,\sigma}$ is the on-site energy of an electron at the site
($i,j$) of spin $\sigma$ ($\uparrow,\downarrow$). $t_l$ and $t_d$ are the
NNH and SNH integrals, respectively. Due to the presence of magnetic flux
$\phi$ (measured in unit of the elementary flux quantum $\phi_0=ch/e$), a
phase factor $\theta_l=2\pi\phi/N$ appears in the Hamiltonian when an
electron hops longitudinally from one site to its neighboring site,
and accordingly, a negative sign comes when the electron hops in the
reverse direction. $\theta_d$ is the associated phase factor for the
diagonal motion of an electron between two neighboring concentric rings.
No phase factor appears when an electron moves along the vertical direction
which is set by proper choice of the gauge for the vector potential
$\vec{A}$ associated with the magnetic field $\vec{B}$, and this choice
makes the phase factors ($\theta_l$, $\theta_d$) identical to each other
for the longitudinal and diagonal motions. Since the magnetic
field corresponding to the AB flux $\phi$ does not penetrate anywhere of
the surface of the cylinder, we ignore Zeeman term in the above
tight-binding Hamiltonian (Eq.~\ref{equcyl}). $c_{i,j,\sigma}^{\dagger}$
and $c_{i,j,\sigma}$ are the creation and annihilation operators,
respectively, of an electron at the site ($i,j$) with spin $\sigma$.
$U$ is the on-site Hubbard interaction term.

\subsection{Theoretical formulation}

To calculate energy eigenvalues, persistent current and related issues 
here we follow exactly the same prescription which we illustrate in the 
earlier section (Sec. II) i.e., Hartree-Fock mean-field approach.

\subsection{Numerical results and discussion}

Throughout the numerical analysis, in this sub-section, we set the 
nearest-neighbor hopping strength $t_l=-1$ and fix $M=2$ i.e., cylinders 
with two identical rings. Energy scale is measured in the unit of $t_l$.
We describe the results in three different parts. In the first part, we 
consider perfect cylinders with only nearest-neighbor hopping integral. 
In the second part, disordered cylinders described with only NNH integral 
are considered. Finally, in the third part we discuss the effect of 
second-neighbor hopping (SNH) integral on the enhancement of persistent 
current in disordered cylinders.

\subsubsection{Perfect cylinders with NNH integral}

For perfect cylinders we choose $\epsilon_{i,j,\uparrow}=
\epsilon_{i,j,\downarrow}=0$ for all ($i,j$). Since here we consider the
cylinders described with NNH integral only, the second-neighbor hopping
strength $t_d$ is fixed to zero.

\begin{figure}[ht]
{\centering \resizebox*{7.5cm}{7cm}
{\includegraphics{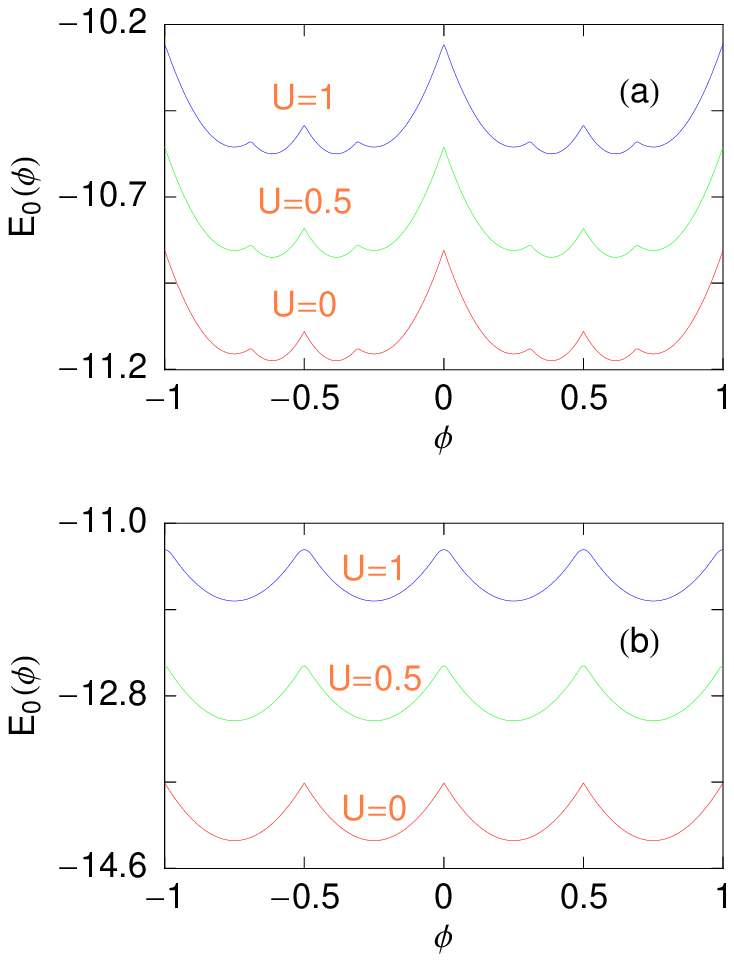}}\par}
\caption{(Color online). Ground state energy levels as a function of
flux $\phi$ for some perfect cylinders with $N=5$ and $M=2$. The red,
green and blue curves correspond to $U=0$, $0.5$ and $1$, respectively.
(a) Quarter-filled case and (b) Half-filled case.}
\label{cylinderenergy1}
\end{figure}

\begin{figure}[ht]
{\centering \resizebox*{7.5cm}{7cm}
{\includegraphics{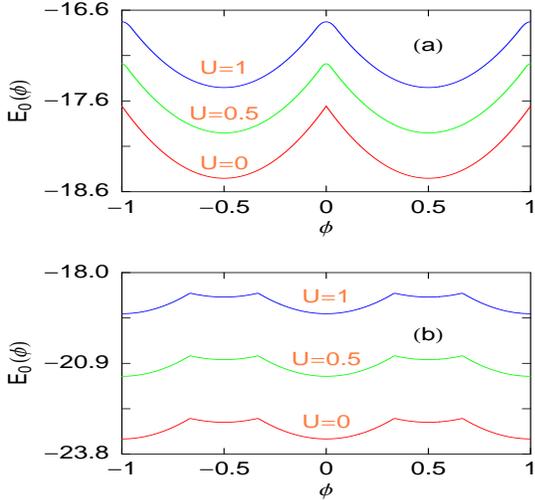}}\par}
\caption{(Color online). Ground state energy levels as a function of
flux $\phi$ for some perfect cylinders considering $N=8$ and $M=2$. The
red, green and blue curves correspond to $U=0$, $0.5$ and $1$, respectively.
(a) Quarter-filled case and (b) Half-filled case.}
\label{cylinderenergy2}
\end{figure}

\vskip 0.2cm
\noindent
{\bf Energy-flux characteristics:}
As illustrative examples, in Fig.~\ref{cylinderenergy1} we show the
variation of ground state energy levels as a function of magnetic flux
$\phi$ for some typical mesoscopic cylinders where $N$ is fixed at $5$
(odd $N$). In (a) the results are given for the quarterly-filled ($N_e=5$)
cylinders, while in (b) the curves correspond to the results for the
half-filled ($N_e=10$) cylinders.
The red, green and blue lines represent the ground state energy levels for
$U=0$, $0.5$ and $1$, respectively. It is observed that the ground state
energy shows oscillatory behavior as a function of $\phi$ and the energy
increases as the electronic correlation strength $U$ gets increased. Most
significantly we see that the ground state energy levels provide two
different types of periodicities depending on the electron filling. At
quarter-filling, ground state energy level gives $\phi_0$ ($=1$, since
$c=e=h=1$ in our chosen unit system) flux-quantum periodicity. On the
other hand, at half-filling it shows $\phi_0/2$ flux-quantum periodicity.
The situation becomes quite different when the total number of atomic sites
$N$ in individual rings is even. For our illustrative purposes in
Fig.~\ref{cylinderenergy2} we plot the lowest energy levels as a function
of $\phi$ for some typical mesoscopic cylinders considering $N=8$ (even $N$).
The curves of different colors correspond to the identical meaning as in
Fig.~\ref{cylinderenergy1}. From the spectra given in
Figs.~\ref{cylinderenergy2}(a) (quarter-filled case) and (b) (half-filled
case) it is clearly observed that the ground state energy levels vary
periodically with AB flux $\phi$ exhibiting only $\phi_0$ flux-quantum
periodicity. Thus it can be emphasized that the appearance of half
flux-quantum periodicity strongly depends on the electron filling as well
as on the oddness and evenness of the total number of atomic sites $N$ in
individual rings. Only for the half-filled cylinders with odd $N$, the
lowest energy level gets $\phi_0/2$ periodicity with flux $\phi$. Now it
is important to note that this half flux-quantum periodicity does not
depend on the width ($M$) of the cylinder and also it is independent of
the Hubbard correlation strength $U$.
Hence, depending on the system size and filling of electrons variable
periodicities are observed in the variation of lowest energy level. It
may provide an important signature in studying magnetic response in
nano-scale loop geometries.

\begin{figure}[ht]
{\centering \resizebox*{7.5cm}{7cm}{\includegraphics{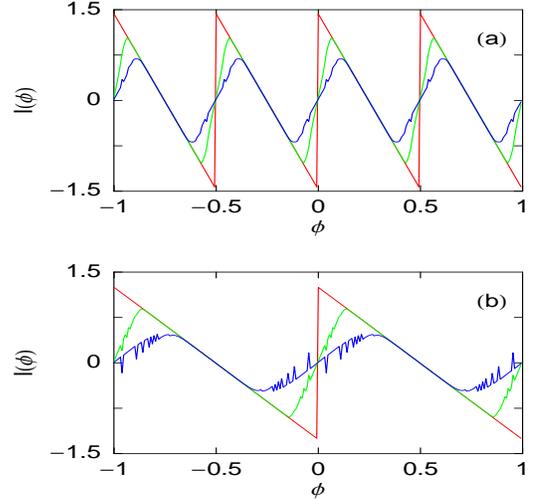}}\par}
\caption{(Color online). Persistent current as a function of flux $\phi$
for some ordered mesoscopic cylinders considering $M=2$. (a) Half-filled
case with $N=15$. The red, green and blue curves correspond to $U=0$, $1.5$
and $2$, respectively. (b) Quarter-filled case with $N=20$. The red, green
and blue curves correspond to $U=0$, $2$ and $3$, respectively.}
\label{cylindercurr}
\end{figure}

\vskip 0.2cm
\noindent
{\bf Current-flux characteristics:}
In Fig.~\ref{cylindercurr} we display the current-flux characteristics
for some impurity free mesoscopic cylinders considering $M=2$. In (a) the
results are given for the half-filled case where we set $N=15$. The red
line corresponds to the current for the non-interacting ($U=0$) case,
while the green and blue lines represent the currents when $U=1.5$ and
$2$, respectively. From the curves we notice that the current amplitude
gradually decreases with the increase of electronic correlation strength
$U$. The reason is that at half-filling each site is occupied by at
least one electron of up spin or down spin, and the placing of a second
electron of opposite spin needs more energy due to the repulsive effect
of $U$. Thus conduction becomes difficult as it requires more energy when
an electron hops from its own site and situates at the neighboring site.
Now both for the non-interacting and interacting cases, current shows
half flux-quantum periodicity as a function a $\phi$ obeying the
energy-flux characteristics since here we choose odd $N$ ($N=15$). The
behavior of the persistent currents for even $N$ is shown in (b) where
we set $N=20$. The currents are drawn for the quarter-filled case i.e.,
$N_e=20$, where the red, green and blue curves correspond to $U=0$,
$2$ and $3$, respectively. The reduction of current amplitude with the
increase of Hubbard interaction strength is also observed for this
quarter-filled case, similar to the case of half-filled as described
earlier. But the point is that at quarter-filling, the reduction of
current amplitude is much smaller compared to the half-filled situation.
This is quite obvious in the sense that at less than half-filling `empty'
lattice sites are available where electrons can hop easily without any cost
of extra energy and the conduction becomes much easier than the half-filled
situation. In this quarter-filled case, persistent currents provide only
$\phi_0$ flux-quantum periodicity following the $E$-$\phi$ diagram. From
these current-flux characteristics it can be concluded that for {\em
`ordered' cylinders current amplitude always decreases with the enhancement
in Hubbard correlation strength $U$.}

\subsubsection{Disordered cylinders with NNH integral}

In order to describe the effect of impurities on electron transport now
we focus our attention on the results of some typical disordered cylinders
described with NNH integral. Here we consider the diagonal disordered
\begin{figure}[ht]
{\centering \resizebox*{7.5cm}{7cm}
{\includegraphics{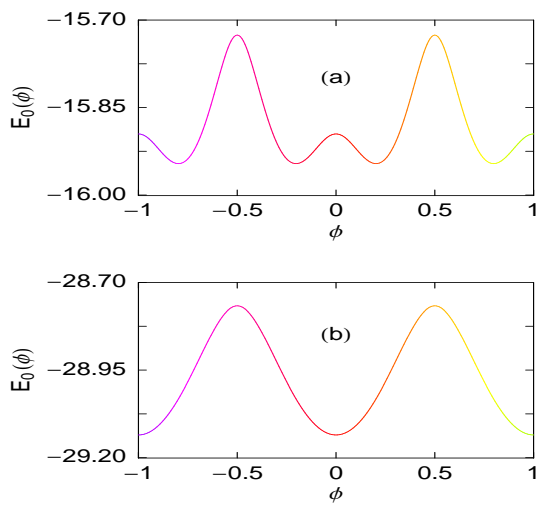}}\par}
\caption{(Color online). Ground state energy level as a function of
flux $\phi$ for half-filled disordered mesoscopic cylinders ($M=2$)
considering $U=1$ and $W=2$. (a) $N=5$ and (b) $N=8$.}
\label{cylinderenergy3}
\end{figure}
cylinders i.e., impurities are introduced only at the site energies
without disturbing the hopping integrals. The site energies in each
concentric ring are chosen from a correlated distribution function
which looks in the form,
\begin{equation}
\epsilon_{j,\uparrow}=\epsilon_{j,\downarrow}=W \cos\left(j \lambda 
\pi\right)
\label{equ7q}
\end{equation}
where, $W$ is the impurity strength. $\lambda$ is an irrational number
and we choose $\lambda=(1+\sqrt{5})/2$, for the sake of our illustration.
Setting $\lambda=0$, we get back the pure system with uniform site energy
$W$. Now, instead of considering site energies from a correlated
distribution function, as mentioned above in Eq.~\ref{equ7q}, we can also
take them randomly from a ``Box" distribution function of width $W$.
But in the later case we have to take the average over a large number of
disordered configurations (from the stand point of statistical average)
and since it is really a difficult task in the aspect of numerical
computation we select the other option. Not only that in the averaging
process several mesoscopic phenomena may disappear. Therefore, the
averaging process is an important issue in low-dimensional systems.

In presence of disorder, energy levels get modified significantly. For our
illustrative purposes in Fig.~\ref{cylinderenergy3} we plot ground state
energy levels as a function of magnetic flux $\phi$ for some disordered
mesoscopic cylinders when they are half-filled. The Hubbard interaction
strength $U$ is set at $1$ and the impurity strength $W$ is fixed to $2$.
In (a) the ground state energy level is shown for a cylinder with $N=5$
(odd), while in (b) it is presented for a cylinder taking $N=8$ (even).
Quite interestingly we see that for the cylinder with odd $N$, the half
flux-quantum periodicity of the lowest energy level disappears in the
presence of impurity and it provides conventional $\phi_0$ periodicity.
Hence, for cylinders with odd $N$, $\phi_0/2$ flux-quantum periodicity
will be observed only when they are free from any impurity. For the
disordered cylinder with even $N$ ($N=8$), the lowest energy level
as usual provides $\phi_0$ periodicity similar to the impurity free
cylinders containing even $N$. Apart from this periodic nature, impurities
play another significant role in the determination of the slope of the
energy levels. The slope of the lowest energy level decreases significantly
compared to the perfect case, and therefore, a prominent change in current
amplitude also takes place.

\begin{figure}[ht]
{\centering \resizebox*{7.75cm}{4.3cm}
{\includegraphics{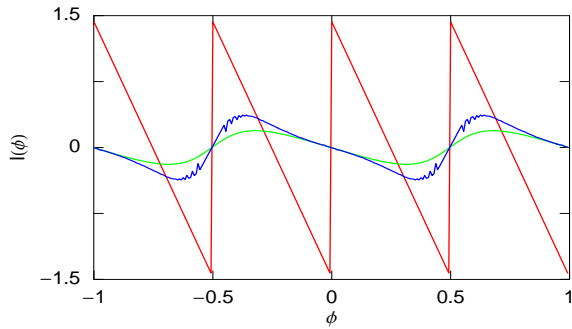}}\par}
\caption{(Color online). Persistent current as a function of flux $\phi$
for a half-filled mesoscopic cylinder considering $N=15$ and $M=2$. The
red line corresponds to the ordered case when $U=0$, whereas the green
and blue lines correspond to the disordered case ($W=2$) when $U=0$ and
$1.5$, respectively.}
\label{cylindercurr1}
\end{figure}

To justify the above facts, in Fig.~\ref{cylindercurr1} we present the
variations of persistent currents with AB flux $\phi$ for a half-filled
mesoscopic cylinder, described in the framework of NNH model, considering
$N=15$ and $M=2$. The red curve represents the current for the ordered
($W=0$) non-interacting ($U=0$) cylinder. It shows saw-tooth like nature
with flux $\phi$ providing $\phi_0/2$ flux-quantum periodicity. The
situation becomes completely different when impurities are introduced in
the cylinder as seen by the other two curves. The green curve represents
the current for the case only when impurities are considered but the
effect of electronic correlation is not taken into account. It shows a
continuous like nature with $\phi_0$ flux-quantum periodicity. The most
important observation is that the current amplitude gets reduced enormously,
even an order of magnitude, compared to the perfect cylinder.
This is due to the localization of the energy eigenstates in the presence
of impurity, which is the so-called Anderson localization. Hence, a large
difference exists in the current amplitudes of an ordered and disordered
non-interacting cylinders and it was the main controversial issue among
the theoretical and experimental predictions. Experimental verifications
suggest that the measured current amplitude is quite comparable to the
theoretical current amplitude obtained in a perfect system. To remove this
controversy, as a first attempt, we include the effect of e-e correlation
in the disordered cylinder described by the NNH model. The result is shown
by the blue curve where $U$ is fixed at $1.5$. It is observed that the
current amplitude gets increased compared to the non-interacting
disordered cylinder, though the increment is too small. Not only that
the enhancement can take place only for small values of $U$, while for
large enough $U$ the current amplitude rather decreases. This phenomenon
can be implemented as follows.
For the non-interacting disordered cylinder the probability of getting two
opposite spin electrons becomes higher at the atomic sites where the site
energies are lower than the other sites since the electrons get pinned at
the lower site energies to minimize the ground state energy, and this
pinning of electrons becomes increased with the rise of impurity strength
$W$. As a result the mobility of electrons and hence the current amplitude
gets reduced with the increase of impurity strength $W$. Now, if we
introduce electronic correlation in the system then it tries to depin
two opposite spin electrons those are situated together due to the Coulomb
repulsion. Therefore, the electronic mobility is enhanced which provides
larger current amplitude. But, for large enough interaction strength, no
electron can able to hop from one site to other at the half-filling
since then each site is occupied either by an up or down spin electron
which does not allow other electron of opposite spin due to the repulsive
term $U$. Accordingly, the current amplitude gradually decreases with $U$.
On the other hand, at less than half-filling though there is some finite
probability to hop an electron from one site to the other available
`empty' site but still it is very small. So, in brief, we can say that
within the nearest-neighbor hopping (NNH) approximation electron-electron
interaction does not provide any significant contribution to enhance
the current amplitude, and hence the controversy regarding the current
amplitude still persists.

\subsubsection{Disordered cylinders with NNH and SNH integrals}

To overcome the existing situation regarding the current amplitude,
in this sub-section, finally we make an attempt by incorporating the
effect of second-neighbor hopping (SNH) integral in addition to the
nearest-neighbor hopping (NNH) integral.

\begin{figure}[ht]
{\centering \resizebox*{7.75cm}{4.3cm}
{\includegraphics{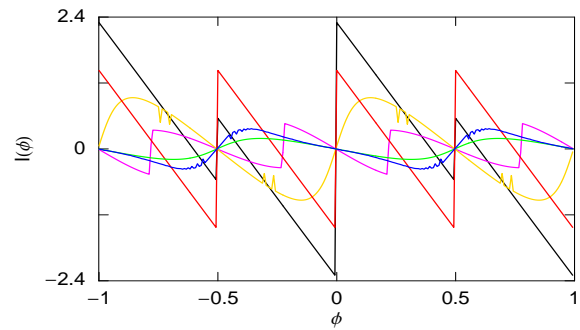}}\par}
\caption{(Color online). Persistent current as a function of flux $\phi$
for a half-filled mesoscopic cylinder taking $N=15$ and $M=2$ in the
presence of NNH and SNH integrals. The black line corresponds to the
ordered case when $U=0$, whereas the magenta and gold lines correspond
to the disordered case ($W=2$) when $U=0$ and $1.5$, respectively. Here
SNH integral is fixed at $-0.6$. The currents shown by the red, green
and blue lines for the ring described with NNH model (identical to
Fig.~\ref{cylindercurr1}) are re-plotted to judge the effect of SNH
integral over NNH model much clearly.}
\label{cylindercurr2}
\end{figure}

A significant change in current amplitude takes place when we include
the contribution of second-neighbor hopping (SNH) integral in addition to
the NNH integral. As representative examples, in Fig.~\ref{cylindercurr2}
we plot the current-flux characteristics for a half-filled mesoscopic
cylinder considering $N=15$ and $M=2$. The black,
magenta and gold lines correspond to the results in the presence of
SNH integral, while the other three colored curves (red, green and
blue) represent the currents in the absence of SNH integral. Here
we choose $t_d=-0.6$. The black curve refers to the persistent current
for the perfect ($W=0$) non-interacting ($U=0$) cylinder and it achieves
much higher amplitude compared to the NNH model (red curve). This
additional contribution comes from the SNH integral since it allows
electrons to hop further. In addition it is also noticed that the current
varies periodically with $\phi$ providing $\phi_0$ flux-quantum periodicity,
instead of $\phi_0/2$ as in the case of NNH integral model (red curve).
Thus, it can be emphasized that $\phi_0/2$ periodicity will be observed
only when the cylinder is (a) free from impurity, (b) half-filled, (c)
made with odd $N$, and (d) described by the nearest-neighbor hopping model.
The main focus of this sub-section is to interpret the combined effect
of SNH integral and electron-electron correlation on the enhancement of
persistent current amplitude in disordered cylinder. To do this first we
narrate the effect of SNH integral in disordered non-interacting cylinder.
The nature of the current for this particular case is shown by the magenta
curve of Fig.~\ref{cylindercurr2}. It shows that the current amplitude
gets reduced compared to the perfect case (black line), which is expected,
but the reduction of the current amplitude is quite small than the NNH
integral model. This is due the fact that the SNH integral tries to
delocalize the electronic
states, and therefore, the mobility of the electrons is enriched. The
situation becomes more interesting when we include the effect of Hubbard
interaction. The behavior of the current in the presence of interaction
\begin{figure}[ht]
{\centering \resizebox*{7.75cm}{4.3cm}
{\includegraphics{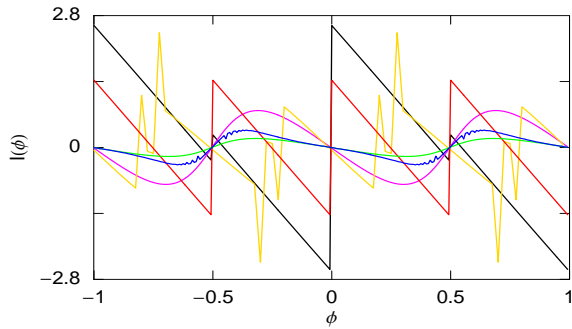}}\par}
\caption{(Color online). Persistent current as a function of flux $\phi$
for a half-filled mesoscopic cylinder taking $N=15$ and $M=2$ in the
presence of NNH and SNH integrals. The black line corresponds to the
ordered case when $U=0$, whereas the magenta and gold lines correspond
to the disordered case ($W=2$) when $U=0$ and $1.5$, respectively. Here
SNH integral is fixed at $-0.8$. The currents shown by the red, green
and blue lines for the ring described with NNH model (identical to
Fig.~\ref{cylindercurr1}) are re-plotted to judge the effect of SNH
integral over NNH model much clearly.}
\label{cylindercurr3}
\end{figure}
is plotted by the gold curve of Fig.~\ref{cylindercurr2} where we fix
$U=1.5$. Very interestingly we see that the current amplitude is enhanced
significantly and quite comparable to that of the perfect cylinder.

For better clarity of the results discussed above, in
Fig.~\ref{cylindercurr3} we also present the similar feature of persistent
current for other hopping strength of SNH integral. Here we set $t_d=-0.8$.
From these curves we see that the current amplitude gets enhanced more as
we increase the SNH strength.

Thus, it can be emphasized that the presence of higher order hopping 
integrals and electron-electron correlation may provide a persistent 
current which can be comparable to the measured current amplitudes.
Throughout the above analysis we set the width of the cylinders at a 
fixed value ($M=2$), for the sake of our illustration. But, all these 
results are also valid for cylinders of larger widths.

\section{A mesoscopic ring with Rashba and Dresselhaus SO interactions}

Finally, in this section we address the magneto-transport properties in
a mesoscopic ring, threaded by an AB flux $\phi$, in the presence of
Rashba and Dresselhaus SO interactions. We establish that the presence of
SO interaction, in general, leads to an enhanced amplitude of the persistent
current. This is another one approach through which we can justify the 
appearance of larger current amplitude in a disordered ring. For this 
discussion we neglect the effect of electron-electron interaction.

\subsection{Model, TB Hamiltonian and the theoretical formulation}

\begin{figure}[ht]
{\centering \resizebox*{6.5cm}{2.5cm}{\includegraphics{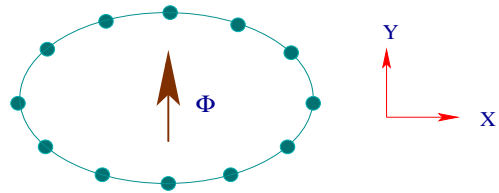}}\par}
\caption{(Color online). A mesoscopic ring threaded by an AB flux $\phi$.}
\label{ringnew}
\end{figure}

The schematic view of a mesoscopic ring subjected to an AB flux $\phi$ 
(measured in unit of the elementary flux quantum $\phi_0=ch/e$) is shown in
Fig.~\ref{ringnew}.
Within a TB framework the Hamiltonian for such an
$N$-site ring is~\cite{sheng,splett,moca} (and references therein),
\begin{equation}
H = H_0 + H_{so}.
\label{equso1}
\end{equation}
Here,
\begin{equation}
H_0 = \sum_n \mbox{\boldmath $c_n^{\dagger} \epsilon_0 c_n$} +
\sum_n \left(\mbox{\boldmath $c_n^{\dagger} t$} \,e^{i \theta} \mbox 
{\boldmath $c_{n+1}$} + h.c. \right)
\label{equso22} 
\end{equation}
and,
\begin{equation}
H_{so}= -\sum_n \left[\mbox{\boldmath $c_n^{\dag} t_{so}$} e^{i\theta} 
\mbox{\boldmath $c_{n+1}^{\dag}$} + h.c. \right]
\label{equso3}
\end{equation}
where,
\begin{eqnarray}
\mbox{\boldmath$t_{so}$} & = & i t_{Rso}\left(\mbox{\boldmath$\sigma_x$} 
\cos\varphi_{n,n+1} + \mbox{\boldmath$\sigma_y$} \sin\varphi_{n,n+1}
\right) \nonumber \\
& - & i t_{Dso} \left(\mbox{\boldmath$\sigma_y$} \cos\varphi_{n,n+1} 
+ \mbox{\boldmath$\sigma_x$} \sin\varphi_{n,n+1}\right).
\label{eqso44}
\end{eqnarray}
$n=1$, $2$, $\dots$, $N$ is the site index along the azimuthal direction
$\varphi$ of the ring. The other factors in Eqs.~\ref{equso22} and 
\ref{eqso44} are as follows.
\vskip 0.2cm
\noindent
\mbox{\boldmath $c_n$}=$\left(\begin{array}{c}
c_{n \uparrow} \\
c_{n \downarrow}\end{array}\right);$
\mbox{\boldmath $\epsilon_0$}=$\left(\begin{array}{cc}
\epsilon_0 & 0 \\
0 & \epsilon_0 \end{array}\right);$
\mbox{\boldmath $t$}=$t\left(\begin{array}{cc}
1 & 0 \\
0 & 1 \end{array}\right)$. \\
~\\
\noindent
Here $\epsilon_0$ is the site energy of each atomic site of the ring.
$t$ is the nearest-neighbor hopping integral and $\theta=2\pi \phi/N$
is the phase factor due to the AB flux $\phi$ threaded by the ring.
$t_{Rso}$ and $t_{Dso}$ are the isotropic nearest-neighbor transfer
integrals which measure the strengths of Rashba and Dresselhaus SO
couplings, respectively, and $\varphi_{n,n+1}=\left(\varphi_n+
\varphi_{n+1}\right)/2$, where $\varphi_n=2\pi(n-1)/N$.
\mbox{\boldmath $\sigma_x$} and \mbox{\boldmath $\sigma_y$} are the
Pauli spin matrices. $c_{n \sigma}^{\dagger}$ ($c_{n \sigma}$) is the
creation (annihilation) operator of an electron at the site $n$ with
spin $\sigma$ ($\uparrow,\downarrow$). Throughout the numerical analysis,
in this sub-section, we choose $t=1$ and measure the SO coupling strength
in unit of $t$.

At absolute zero temperature ($T=0\,$K), the persistent current in the
ring described with fixed number of electrons $N_e$ is determined by,
\begin{equation}
I\left(\phi\right)=-c\frac{\partial E_0(\phi)}{\partial \phi}
\label{equso2}
\end{equation}
where, $E_0(\phi)$ is the ground state energy. We compute this quantity
by exactly diagonalizing the TB Hamiltonian (Eq.~\ref{equso1}) to 
understand unambiguously the role of the RSOI interaction alone on 
persistent current.

\subsection{Numerical results and discussion}

\vskip 0.2cm
\noindent
{\bf Energy-flux characteristics:}
Before presenting the results for $I(\phi)$, to make the present
communication a self contained study, we first take a look at the energy
\begin{figure}[ht]
{\centering \resizebox*{8cm}{8.2cm}{\includegraphics{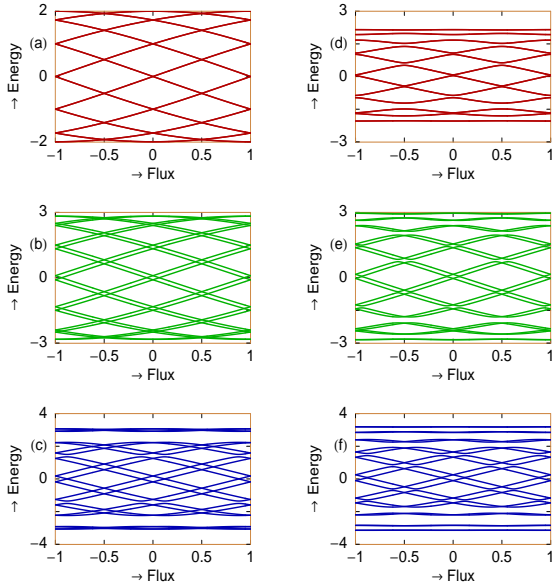}}\par}
\caption{(Color online). $E$-$\phi$ curves of a $12$-site ring, where the
$1$st and $2$nd columns correspond to the results for the ordered ($W=0$)
and disordered ($W=1$) cases, respectively. The red, green and blue lines
correspond to $t_{Rso}=t_{Dso}=0$; $t_{Rso}=1$, $t_{Dso}=0$ and $t_{Rso}=1$,
$t_{Dso}=0.5$, respectively.}
\label{energy}
\end{figure}
spectrum of both an ordered and a disordered ring with and without the
SO interactions, as the flux through the ring is varied. 
In Fig.~\ref{energy} the flux dependent spectra are shown for a $12$-site
ordered ring and a randomly disordered one (with diagonal disorder) in the
left and the right columns respectively. Clearly, disorder destroys the
band crossings observed in the ordered case. The presence of the RSOI and
the DSOI also lifts the degeneracy and opens up gaps towards the edges
of the spectrum.

\vskip 0.2cm
\noindent
{\bf Enhancement of persistent current:}

\begin{figure}[ht]
{\centering \resizebox*{7.5cm}{4cm}
{\includegraphics{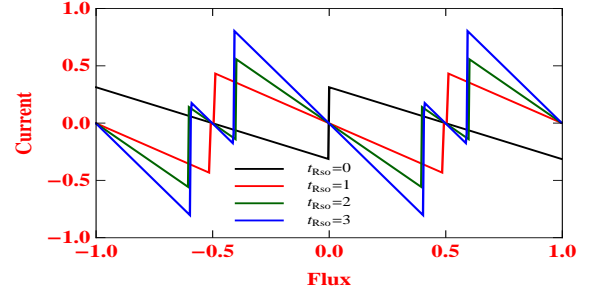}}\par}
\caption{(Color online). Current-flux characteristics of a $80$-site
ordered ($W=0$) half-filled ring for different values of $t_{Rso}$ when
$t_{Dso}$ is set at $0$.}
\label{current}
\end{figure}

\vskip 0.1cm
\noindent
$\bullet$ {\underline{An ordered ring:}} In Fig.~\ref{current} we
examine the effect of the RSOI on the persistent current of an ordered
ring with $80$ sites. The DSOI is set equal to zero. We have examined
both the non-half-filled and half-filled band cases, but present
results for the latter only to save space. With increasing strength
of the RSOI the persistent current exhibits a trend of an increase in
its amplitude. Local phase reversals take place together with the
appearance of kinks in the current-flux diagrams which are however,
not unexpected even without the RSOI, and are results of the band
crossings observed in the spectra of such rings. The amplitude of
the persistent current at a specific value of the magnetic flux is of
course not predictable in any simple manner, and is found to be highly
\begin{figure}[ht]
{\centering \resizebox*{7.5cm}{4cm}
{\includegraphics{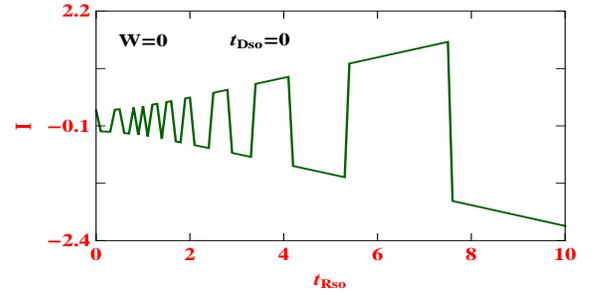}}\par}
\caption{(Color online). Persistent current at a particular AB flux
($\phi=0.25$) as a function Rashba SO interaction strength for an ordered
($W=0$) half-filled ring with $N=60$ when $t_{Dso}$ is set to zero.}
\label{typicurr3}
\end{figure}
sensitive to the number of electrons $N_e$ (i.e., the filling factor).
Issues related to the dependence of the persistent current on the
filling factor have been elaborately discussed by Splettstoesser
{\em et al.}~\cite{splett}.

The persistent current in an ordered ring also exhibits interesting
oscillations in its amplitude as the RSOI is varied keeping the magnetic
flux fixed at a particular value. The oscillations persist irrespective
of the band-filling factor $N_e$, with or without the presence of the
DSOI. In Fig.~\ref{typicurr3} the oscillating nature
of persistent current is presented for a $60$-site ordered ring in the
half-filled band case when $\phi$ is set at $\phi_0/4$. The current
exhibits oscillations with growing amplitude as the strength of the
RSOI is increased.

\begin{figure}[ht]
{\centering \resizebox*{7.5cm}{4cm}
{\includegraphics{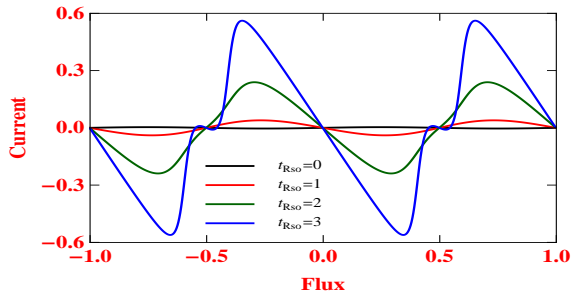}}\par}
\caption{(Color online). Current-flux characteristics of a $80$-site
disordered ($W=2$) half-filled ring for different values of $t_{Rso}$
when $t_{Dso}$ is fixed at $0$.}
\label{currentdisorder1}
\end{figure}

\vskip 0.1cm
\noindent
$\bullet$ {\underline {A disordered ring:}} We now present the results
for a disordered ring of $80$ sites in Fig.~\ref{currentdisorder1}.
Disorder is introduced via a random distribution (width $W =2$) of the
values of the on-site potentials (diagonal disorder), and results averaged
over sixty disorder configurations have been presented. The DSOI remains
zero. Without any spin-orbit interaction, disorder completely suppresses
the persistent current (an effect of the localization of the electronic
states in the ring), as it is observed in Fig.~\ref{currentdisorder1}
(black curve). With the introduction of the RSOI, the current starts
increasing, and for $t_{Rso}=3$ (blue curve), increases significantly,
\begin{figure}[ht]
{\centering \resizebox*{7.5cm}{4cm}
{\includegraphics{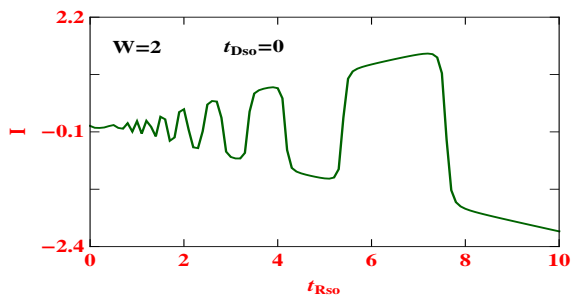}}\par}
\caption{(Color online). Persistent current at a particular AB flux
($\phi=0.25$) as a function Rashba SO interaction strength for a
disordered ($W=2$) half-filled ring with $N=60$ when $t_{Dso}$ is set
to zero.}
\label{typicalcurrentdisorder}
\end{figure}
attaining a magnitude comparable to that in a perfectly ordered ring.
It is to be noted that the strength of the RSOI is strongly dependent
on gate voltage. An enhancement
of the persistent current in the presence of disorder can be achieved
even with much lower values of the RSOI parameter compared to what have
already been presented in the figures. To achieve this one needs
to increase the size of the mesoscopic ring. We have checked this with
a $100$-site ring where even with $t_{Rso}=0.5$ the current increases
by an order of magnitude compared to the case when $t_{Rso}=0$. However,
we present the results using a somewhat larger values of $t_{Rso}$ for a
better viewing of the results. Similar observations are made by setting
$t_{Rso}=0$ and varying $t_{Dso}$.

Disorder introduces quantum interference which leads to localization of
the electronic states. RSOI, on the other hand, introduces spin flip
scattering in the system,
which can destroy quantum interference effect, leading to a possible
delocalization of the electronic states. This leads to an enhancement
of the persistent current in the presence of disorder. The competition
between the strength of disorder and the RSOI is also apparent in
Fig.~\ref{typicalcurrentdisorder}. For small values of the RSOI, the
disorder dominates. As the strength of the RSOI is increased, the spin
flip scattering starts dominating over the quantum interference effect,
and finally the oscillations become quite similar to that in a ballistic
ring. As the SO interaction is a natural interaction for a quantum ring
grafted at a heterojunction, we are thus tempted to propose that the
spin-orbit interaction is responsible for an enhanced persistent current
in such mesoscopic disordered rings.

Before we end this section, we would like to mention that the presence 
of DSOI alone leads to exactly similar results as expected, since the 
Rashba and the Dresselhaus Hamiltonians are related by a unitary
transformation. This does not change the physics. We also examine the
behavior of persistent current in presence of both the interactions. 
The amplitude of the current does not increase significantly compared 
to the case where only one interaction is present. However, the precise 
magnitude of the current is sensitive to the strength of the magnetic 
flux threading the ring. The observation remains valid even when the 
strengths of the RSOI and DSOI are the same.

\section{Summary and Conclusions}

In this review we have demonstrated the magneto-transport properties in 
single-channel rings and multi-channel cylinders based on the tight-binding 
framework. Several anomalies between the theoretical and experimental 
results have been pointed out and we have tried to remove some of these
discrepancies. The main controversy is associated with the proper 
determination of persistent current amplitude. We have addressed two 
different possibilities of getting enhanced current amplitudes and proved 
that the currents are quite comparable to the experimentally predicted 
results. In one approach
we have concluded after an exhaustive numerical calculation that the 
presence of higher order hopping integrals and electron-electron interaction
can provide a persistent current which may be comparable to the actual
measured values. We have justified these results both for the single-channel
and multi-channel cases. In other approach we have established that in
presence of spin-orbit interaction a considerable enhancement of current
amplitude takes place, even for the non-interacting nearest-neighbor hoping
model, and the magnitude of the current in a disordered ring becomes almost
comparable to that of an ordered one. In addition to these, we have also
analyzed the detailed band structures, low-field magnetic susceptibilities,
variation of electronic mobility and some other issues to get the full
picture of the phenomena at the meso-scale and nano-scale levels.

Here we have considered several important approximations by ignoring the 
effects of temperature, electron-phonon interaction, etc. Due to these 
factors, any scattering process that appears in the systems would have 
influence on electronic phases. At the end, we would like to say that we 
need further study in such systems by incorporating all these effects.

\vskip 0.2cm
\noindent
{\bf Future directions and opportunities:}
Although the studies involving the mesoscopic rings and cylinders have 
already generated a wealth of literature there is still need to look 
deeper into the problems both from the point of view of fundamental 
physics and to resolve a few issues that have not yet been answered in 
an uncontroversial manner. For example, it may be interesting to study 
the combined effect of electron-electron interaction and spin-orbit
interaction on magneto-transport properties. Specially, the effect of
SO interaction on electron transport in mesoscopic and nano-scale 
semiconductor structures should be carefully examined. The principal 
reason is its potential application in spintronics, where the possibility 
of manipulating and controlling the spin of the electron rather than its
charge, plays the all important role~\cite{zutic,ding,bellucci,san9}.

\end{document}